2022.08.18

# Interviews about modern astrometry

*by Erik Høg – Niels Bohr Institute - Copenhagen*

Michael Perryman has interviewed some of the scientists and project leaders in the Hipparcos and Gaia missions, the interviews with photos of the persons are given at his site: https://www.michaelperryman.co.uk .

Michael has also written essays - 85 to date ! - about results from the Gaia mission and they are placed at his site.

Three of the interviews are with me and transcriptions, co-authored with Michael, are provided below with the titles:

**1. An interview about astronomy and astrometry up to 1980**

**2. An interview about the revival of astrometry after 1980**

**3. The billion-star astrometry after 1990**

The third interview begins in 1990 when I had the first ideas for a *Hipparcos successor*. In 1992 I made a detailed design with direct imaging on CCD detectors in a satellite proposal called Roemer. In 1993 a supposedly better option was proposed with the acronym GAIA where the capital "I" stood for Interferometer. In 1998, however, interferometry was shown to be unsuited for the purpose and we returned to the original idea from 1992 for the further development. The name was later changed to Gaia - for the sake of continuity.

*62 pages = 1+22+19+20, including abstracts and references.*

2021.08.26

# An interview about astronomy and astrometry up to 1980

*by Michael Perryman and Erik Høg*

ABSTRACT: Michael Perryman invited Erik Høg to this interview in July 2021. Most surprising for a young astronomer is perhaps that space astrometry nearly failed, but is so famous today from the great impact on astronomy by the Gaia satellite measuring two billion stars. But the approval in 1980[1] of the first astrometry satellite Hipparcos was in fact reached with a very narrow margin. The lay reader may enjoy the first few pages about a schoolboy in a Danish village building his own telescopes in the 1940s in order to observe variable stars in a wonderfully dark sky. The technology may then stop her/him from further reading so that the wise words would be missed which were spoken at a meeting in October 1975, just one sentence that opened the door for the ensuing revolution of astronomy. In the interview, Erik Høg again emphasizes the crucial contributions to space astrometry by Pierre Lacroute, Jean Kovalevsky and Lennart Lindegren especially in the 1970s: There were simply nobody around who could have played their roles, without them there would have been no Hipparcos approved in 1980, and probably never. This report contains the questions by the interviewer, the answers by Erik Høg, a link to the audio recording and references.

---

Here follows a script of the recording. It quotes Erik closely but Michael not so closely, with slanted text. The audio is available at this link: http://www.astro.ku.dk/~erik/xx/erik-hoeg-final.mp3 ,
but you should rather open it on Michael's website with photos and other interviews: https://www.michaelperryman.co.uk/gaia-project-interviews

The minutes and seconds are tagged at tm=.... Total duration 48 minutes, speaking takes 1 minute per 130 words.

**Introduction**

Michael Perryman:

---

[1] The approval was in fact in **1980**, not in 1981 as said in the interview.

*Gaia is a scientific satellite of the European Space Agency launched in 2013. Today it is still thoroughly scanning the dark sky from a deep space orbit more than a million km from Earth. It is in the process of constructing a remarkable three-dimensional survey of more than two billion stars throughout our galaxy. I'm Michael Perryman, and in a series of conversations I am talking to scientists involved in different aspects of the mission.*

*Joining me today is someone I worked closely with for more than 30 years: Erik Høg, Professor Emeritus of the Niels Bohr Institute, University of Copenhagen.*

*Erik was one of the people responsible for getting the Hipparcos space astrometry mission adopted during the 1970s he was one of the scientific leaders of Hipparcos in the 1980s and 1990s and he was central in the developments of Gaia in the 1990s and 2000s but his innovative contributions to astrometry pre-dated even the first ideas of Hipparcos and his work continues today into a future beyond Gaia.*

*In this conversation, we're covering the time up to the acceptance of Hipparcos in 1981, in other words ground-based astrometry in the second half of the 20th century.*

*What was ground-based astrometry like in the decades before Hipparcos How did Hipparcos develop over the years before its acceptance by ESA in 1981? And what were the challenges in getting this pioneering mission accepted by European scientists and policy makers?*

*Let's find out!*

*Erik Høg: a very warm welcome.*

Hi, I am grateful to you Michael for inviting me to this interview, I recall with pleasure our 25 years of collaboration when you led the science teams of Hipparcos and Gaia.

**Background**

*Erik: you were born in 1932, and you have recently celebrated your 89th birthday. What set you off on your long career as an astronomer?*

Well, as a boy, I read a lot of books about physics, mathematics and astronomy. One morning I saw Venus in the telescope the teacher had set up at the school. The teacher lent me a small telescope for a month. But after that, in 1946 [1] when I was 14 I realized I had to built my own telescopes. I first built one from two spectacle lenses, then I ground and polished the mirrors for two telescopes and observed variable stars. The telescope mounting was built by the village blacksmith.

My teachers now and later, helped and encouraged me in my interests. I must say that most people I ever met in life were nice and helpful. I could read and write before I started with six years in the village school. That was in Frejlev on the island Lolland. On the first day in school, the teacher asked how far we could count. I answered that I could go on counting because I could always say the number that followed: so in effect I knew there are infinitely many numbers. I should add that where I lived there was a completely dark sky in the country side. So I could see the stars and that was wonderful.

tm=04:00

*You were a student at Copenhagen University between 1950-1956. What were your undergraduate studies?*

In 1950 I moved to Copenhagen which is 100 km away from where I had lived with my family where nobody had any further education than technical school. I wanted to become a teacher because my teacher in the gymnasium had said I would not be able to live from astronomy. Therefore I studied mathematics, chemistry, physics and astronomy for three years and then specialized in astronomy. I was well received at the Copenhagen Observatory where we were only two or three students. My tutor was Peter Naur just four years my senior and very bright in both astrometry and astrophysics. He was very engaged in computer development, I learnt much about computers and electronics from him.

In 1953 a new meridian circle was set up at a new site at the village Brorfelde 50 km from Copenhagen and I was set to measure its stability [1]. I could buy a small motorbike to go there for observing. Sometimes I slept in a haystack when clouds came because there were no houses where I could stay. But the fine instrument fascinated me and I found astrometry very interesting and very important for astronomy. Astrometry gives astronomers the positions and motions of stars and also the distances.

*What was really driving you at this time?*

Just to do what I had been asked to do, to test the stability. And then I discovered this fantastic instrument, the mechanics interested me, and then the thoughts about the importance of astrometry.

What drives me in life is to do something useful, something needed, especially if I see that nobody else will do it, then I have to do if I possibly can! The telescopes I built as a school boy were needed for my observations so I had to build them myself. That was probably my tacit natural logic.

tm=06:27

*What came after your undergraduate studies?*

After completing the studies I served as a conscript soldier for a year. I worked in a laboratory where we used electronic counters [1 Sect.4]. Then I worked with Peter Naur near the meridian circle at Brorfelde where houses had now been built. But in 1957 Naur said I should go abroad to another observatory, but I said that I was not ready, I had to learn more before going. Then he said you have to go now, it is out there that you will learn!

*You've also talked about the influence of Bengt Strömgren?*

Yes, Strömgren was director of the Copenhagen Observatory but he spent most of the time in the USA where he was director of two big observatories. So I only saw him in the summer time in Copenhagen, but I corresponded with him about my work with the meridian circle.

Strömgren was one of the great astronomers of the 20th century, mainly an astrophysicist, but very keen on promoting astrometry. He pioneered photoelectric astrometry by an experiment in 1925 which triggered my invention of photon counting astrometry in 1960. He ordered a new meridian circle in 1940 which should become the main instrument of the new observatory, and he supported very actively the approval of Hipparcos in 1980.

tm=08:14

*You then moved on to spend 15 years, from 1958-1973, at the Hamburg Observatory in Northern Germany. What took you from Copenhagen to Germany?*

One of the reasons for choosing the Hamburg Observatory was that its director Otto Heckmann had visited Copenhagen in 1955 and he had initiated several meetings [11] of astronomers from Germany, Denmark, Sweden and Norway.

I came to Hamburg on a 10 month fellowship from Deutsche Akademische Austauschdient. This was extended by a NATO Science Fellowship helped by a friend from Copenhagen who worked on the programme in Paris. I had no superior in Hamburg. I once asked Otto Heckmann for permission to travel, he answered that I was completely free, could do what I wanted. Of course we always spoke German and I am still fluent.

They appreciated my work and I became civil servant, staying on for a total of 15 years and I am now enjoying a monthly pension from Germany.

When I had worked there for about three years, I remember a colleague telling that they were trying to get a permanent position for me. I did not believe him. In fact I was never thinking of a career but only about doing my important work - and also about finding a good wife to live with for the rest of my life. And I succeeded by finding her in Denmark in 1961, but that is another story. Our three children were born in Germany.

I never thought of becoming a professor, a professor was somebody like Strömgren, and I could never match him.

tm=10:18

*How large was the Hamburg Observatory at that time?*

Hamburg was perhaps the largest observatory in Germany at that time. There was perhaps a staff of 50 people including the technical staff and about ten scientists and ten research students.

*What were the big questions in astronomy that were being discussed and funded at this time?*

One of the biggest questions in astronomy in the 1960s was about the size of the universe, since Walter Baade had just found that all distances to galaxies should be doubled. Baade had worked in Hamburg before moving to California.

People in Hamburg measured photometry of many open clusters on photographic plates taken with the big Schmidt telescope.

*There was no internet, nor indeed computers as we know them today. Presumably you read about advances in journals, but only through paper publications and preprints?*

Yes, and there were weekly seminars on the Saturday where everyone was expected to attend in the beautiful library.

*There was, presumably, little international travel?*

I had visited the big international conferences in 1958 in Moscow and in 1961 in California, financed from Denmark, much helped by Einar Hertzsprung who lived in Denmark. We from Hamburg visited e.g. Jena, Tübingen and Göttingen for common meetings.

tm=12:20

*You mentioned Einar Hertzsprung, a very big name in astronomy someone who was active around a hundred years ago. Did you meet him?*

Yes, I have met him many times and visited him many times where he lived which was near the railway station of Tølløse where you went if you should go to the new observatory site at Brorfelde. He invited me to come there and paid my ticket and accompanied me back to the train. We talked and sometimes I measured a plate on his machine and he always ended the talks by inviting me to the other room where we would have a dessert served by his daughter.

*What excited you in science: observing, or instrumentation, or finding out how the Universe worked?*

I often visited the people working in the big Schmidt. I saw they noted the photometer reading on paper and later punched the numbers on cards to be processed in the IBM 650 computer in the city of Hamburg which was 20 km away. I said the reading could be punched directly on cards. This was considered to be a good idea, so I designed and built the digital electronics.

In those years I had to build, e.g., even the power supply from basic components. I also digitized a spectrum scanner in order to measure low-dispersion spectra taken with the Schmidt.

So besides of digitizing I was working on an astrophysical project, but in 1960 I got a very good idea for astrometry, an idea that would lead to the satellite Hipparcos.

**Contributions before Hipparcos**

tm=14:16

*Your time at Hamburg was before there was any discussions of a space mission. I looked up your very first paper, published in 1960 in the Annals of the Hamburg Observatory: "Proposal for a Photoelectric Meridian Circle". Your central idea was that starlight from the telescope would pass through a fixed grating and be detected by a photomultiplier*

*tube. Indeed the first figure in that paper looks rather rather like the principles that would be used two decades later for Hipparcos . Why was this such a useful advance?*

Yes, you describe my idea quite nicely. The idea came to me on the 22 July 1960 [2]. My notes say that I worked up to 12 hours per day on that subject during the following time. I set out to study all aspects of the idea: optimal design, technical feasibility, theoretical basis and limitations e.g. due to the Earth atmosphere, the expected accuracy and I published a number of papers during the following years. I presented it at a conference in the autumn of 1960 in Jena.

I have had the luck often in life to get good ideas at the right place at the right moment of time. In this case the idea was not born out of any immediate local need, but because the technical possibilities for digital astrometry had come together in my mind. The idea was born from Strömgren's experiments with slits in 1925, from Naur's ideas about computers, and from what I had seen at the big technical fair in Hannover.

But Otto Heckmann immediately saw the potential for the planned expedition to Perth in Western Australia with the Hamburg meridian circle. This expedition was part of international undertaking by ten observatories. They should observe 20,000 stars of SRS, it was a Southern Reference Star catalogue.

They had planned to use the old visual micrometers as they had just used to observe reference stars in the northern sky. But with my digital method all observations could be punched on tape and be treated in a dedicated computer soon after.

But Heckmann must have asked himself: can I trust that the young Dane, Erik Høg will stay on for the long development? He asked me in a nice way and I answered: "this idea is worth several years of my life". It became a success, and Heckmann was very soon able to get the funding in place.

tm=17:30

*This instrument idea was used on the Hamburg Meridian Circle between 1960-1967. Was it a success?*

It became a success after a long struggle. In fact it was NOT USED in Hamburg in the time you mentioned. It was only tested there because the development itself took six years, much longer than expected. It became a success during the five years in Perth from 1967-72, thanks to a very dedicated and capable team led by the experienced Dr. Johann von der Heide and the young technical assistant Gerhard Holst.

There were up to 10 observers. The instrument was really only a prototype and met with many problems.

*I read in your 1960 paper that the results were recorded on punched tape and in fact I remember punched tape as a recording mechanism from my own PhD in radio astronomy in the late 1970s. Tell us a little more how the observations were actually made in practice.*

Observation at the instrument was done by two people, usually a married couple. The wife was sitting with the star list in a small cabin at the instrument. She set the star number on toggle switches and told her husband the declination of the star to be observed. He turned the telescope as needed and started the punching on tapes when he saw the star entering the field of view. So the instrument was in fact the first semi-automatic meridian circle, this was a revolution. It produced about 10 big rolls of tape every night.

They were processed on the dedicated computer by the same observing team the next day they were on duty. This computer was a GIER [3] of Danish manufacture. GIER was one of the first transistorized computers and it was crucial for the project because it was ten times faster than the cheapest IBM computer we could have afforded, but of course any laptop today is a million times faster than GIER. I had written all the programs in ALGOL 60, which was the new language to programming in which my former tutor Peter Naur had a leading role. He won the 2005 Turing-award, also known as the "Nobel-prize of computing science."

*Another big chapter in your early work was the Perth 70 Catalogue of about 25,000 stars. Tell us more about that.*

Yes, I visited several times for a couple of weeks to help solve the problems. The result was the Perth 70 Catalogue of about 25,000 stars, published in 1976, in the Annals of the Hamburg Observatory. Much later I published several long historical reports [2] about the work, mostly in Nuncius Hamburgensis, a series of books edited by Dr. Gudrun Wolfschmidt. I am very grateful for her frequent encouragements to write and to speak at her meetings on the history of astronomy during the recent ten years.

*I think you formed strong and long-lasting relations with the German astronomical community at that time, which went on to be important in your work on Hipparcos and Gaia. Who were your main contacts then and in later years in Germany?*

I knew practically all astronomers in Germany, Walter Fricke was one of them, Hans Siedentopf in Tübingen another, the list is too long to continue, but I enjoyed the time in Germany and I still have contact to many.

**The Hipparcos years**

*You moved back from Hamburg to Copenhagen to take up a post at the Copenhagen University Observatory in 1973, where you remained until your retirement in 2002, when you moved to the Niels Bohr Institute as guest professor. Everything we've talked about so far took place before the ideas of a space mission came about. But let's turn to that. The first ideas for a space mission were put forward by Strasbourg astronomer Pierre Lacroute at the thirteenth General Assembly of the IAU in Prague in 1967. When did you hear about these ideas for the first time?*

Lacroute presented his ideas of space astrometry [6] [10] at a meeting in Bordeaux in July 1965 [4]. This was the first time that such type of astronomy was proposed for a space mission. The potential advantages were clear: no atmosphere and no gravity, and perhaps thermal stability if that would be technically feasible.

*How did you react when you saw the ideas of Lacroute set out on paper? Did you think that it was feasible?*

I attended the presentation by Lacroute in Prague in 1967, but to me and most others the technical problems seemed utterly underestimated. Also because he said it could be done for 10 million French franc.

*Who else in Europe was getting excited by the idea of a space mission for astrometry? Was it mainly astronomers in France?*

Yes, the proposal did not start any activity outside France, people were interested, but it was fortunately shared by other French astronomers. This was really Professor Pierre Lacroute's great vision.

*Was there some specific moment that CNES abandoned this as a purely French - or French-led - mission?*

Yes, I know that in 1970 CNES decided not to pursue any purely French space missions, but only European programs. Especially Jean Kovalevsky was able to convert space astrometry from being a national project to become European, through ESA.

*How did the idea of a space mission become something considered by ESA? Who else was pushing this idea?*

Jean Kovalevsky, he was director of a new astrometric observatory in the Southern France and he was really bringing the idea to ESA. Would such a mission interest a sufficiently important scientific community? That was the question asked, and for this reason Kovalevsky convened a Symposium on Space Astrometry in Frascati. It was held in October 1974 and gathered 41 participants from Europe and the United States.

In the two days Lacroute presented his ideas. There were many other presentations, one of them by me where I predicted that automatic meridian circles could compete with space astrometry as predicted by Lacroute.

*Are you saying that, at this time, you didn't really believe in space astrometry?*

No, I was not at all in favour of space astrometry at that time. I could not have fostered the idea of space astrometry since my ideas for astrometry went in a different direction. I must say that without Pierre Lacroute there would have been no Hipparcos mission approved in 1980.

The same is true for Jean Kovalevsky: So Lacroute was the father of space astrometry and Kovalevsky was the architect, I would say. I knew the scene of astrometry of that time, I knew all the relevant people and I am sure there were nobody who could have played their roles.

*This is a remarkable little part of the story Erik, that someone who would go on to be such a strong and leading advocate of space astrometry should start being so sceptical!*

tm=26:48

*Who were the ESA staff that your first came across, and in what capacity?*

I got a call from ESA in Paris where I was asked if I would join a study group on space astrometry for a meeting. My answer was YES. I did not believe it was a good idea, but if ESA wanted my advice I had to go and give it.

*What happened at that meeting?*

At the first meeting of the group in October 1975, Lacroute presented his two ideas called Spacelab option and TD satellite. I asked the chairman Vittorio Manno which of the two ideas he wanted us to discuss first. He answered very wisely: "Just forget what is on the table, only think of how you can use space technology for your science."

That advice had a great effect on me, Vittorio Manno spoke the right words to turn me towards Hipparcos. I could suddenly think freely. I had never been interested in space technology, but I could go back to Denmark where I was able to make a quite new design of an astrometry satellite which I called TYCHO [7], much simpler and more accurate than Lacoute's options. After six weeks I sent the proposal to ESA where it was well received.

Soon after, I was invited to visit ESTEC for a couple of days. I was tutored by Maurice Schuyer on e.g. low earth orbits. I explained the revolving scanning required for my design and I remember that they could not understand what I meant, and they said that such a scanning motion had never been tried before.

But when Hipparcos had been approved they soon understood everything very well these engineers.

*What were the key features of Lacroute's original design?*

Lacroute's original design had three key features which were kept in Hipparcos: The satellite should rotate and systematically scan the entire sky. The satellite should have a split mirror in front of a telescope so that two fields on the sky one after the other would come to the focal plane. The stars should cross over slits in the focal plane and be detected by photon counting.

But other features in his design were too complicated and unrealistic in my view.

*What were the biggest problems that you foresaw in this early concept?*

It was too inefficient to use only slits and photomultipliers, I therefore introduced a grid and an image dissector tube. This made the light collection 100 times more efficient, so that a smaller telescope would be sufficient. My design with a star mapper was able to use an input catalogue so that the observing program could be limited to 100,000 stars chosen for their scientific interest.

Another important feature was that measurement should be only in one direction, only along the scan, no measurement in the cross direction should be made. Therefore the split mirror could be in only two parts, much simpler to manufacture, and the data reduction would be simpler.

Also a new scanning was proposed: revolving scanning where the spin axis is pointed at a fixed angle to the sun. This required active attitude control.

All these features were new. Lacroute quickly adopted an image dissector tube and only this feature in a new design [7] of a scanning satellite. His design was called Option B in the following years while the TYCHO design was called Option A.

*What happened to Lacroute's design?*

In 1977 our team at ESA decided that an input catalogue should be used as needed in my Option A, which was now often called Astrometry Satellite. This could be decided after alternative options had been excluded.

So, Lacroute's Option B was not much discussed. He considered it to be simpler than Option A, but it was not. Lacroute was not a good instrument maker, but he had the immense courage to try during ten years because he had a great vision for science. And he learnt English at an advanced age for that reason.

*We have been talking about some of the technology design elements. But a key aspect of getting a mission like this adopted is to build the support for Hipparcos throughout the scientific community. You played a big part in building up this support for Hipparcos throughout Europe - please tell us about that.*

Back in Denmark I started an "inquiry on projects" [6 A2.1] after the use of an input catalogue had been decided. I visited a number of astronomical institutes especially in Germany to give talks about the project. I also distributed to astronomers mostly in ESA countries a questionnaire asking for their favourite scientific projects and how many stars they would like to have observed by the astrometry satellite. Catherine Turon did similarly in France. These answers were basis for the Hipparcos Input Catalogue later compiled by the Input Catalogue Consortium.

*You also recognised that the processing of the data on the ground was going to be a big challenge - given the state of computers. How did you and others start to address this?*

Even with my new design the data reduction was a formidable task with the computers available in 1976: it was to derive positions, proper motions and parallaxes for 100 000 stars from 10 million angular measures in the course of three years. But fortunately, I already knew Lennart Lindegren [5] since 1973 when he was a 23 year old student at Lund Observatory, how I met him there is story of its own.

In September 1976 I introduced him to space astrometry and my design. After four weeks he sent me a nicely typed report with all the mathematical formulae for the method which was later used during the mission, the “three-step method”. Two weeks later came a report with the first simulations and expected accuracies.

Without his unfailing genius I would say in all mathematical, computational and optical matters the project would not have been ripe for approval in 1980, and probably never.

*The data reduction was a possible line of attack for the other competing projects during the decision process in 1980?*

Oh yes, and without Lindegren we would have lost.

So, I agreed completely when a colleague said to me: "The best you have ever done for astronomy was to find Lennart." But it was not at all easy to have Lindegren included in the Science Team selected for the Phase A study. The members of this team were selected at a meeting in late 1976 of the ESA Astronomy Working Group. At the end they found that there would be too many members if Lennart were included and it was proposed that I could still collaborate with him since he lived not far from Copenhagen. I said then that it was more important that Lennart became a member and that I could stay outside. We both became members.

tm=35:15

*Indeed, Lennart Lindegren played a crucial part in the planning and execution of Hipparcos in the 1970s, 1980s, and 1990s, as well as in the planning and execution of Gaia, in which he continues to be deeply involved today. This is not the time to go into Lennart's contributions, but can you tell us about how your paths first crossed, and how you went about encouraging him to contribute more to space astrometry?*

In September 1973 I heard from a Swedish astronomer that there was a young student in Lund who was working on the old meridian circle. Soon after I visited Lund to give a seminar and there I was introduced to the young Lennart Lindegren. He showed me modestly the very nice improvements he had made on the old instrument. I said I was impressed, but there was still a long way to an operating instrument. He said he knew that and that he had done the work only because he was impressed by the fine mechanics.

I offered him two subjects for a PhD. Soon after he began work on meridian circle observations of the major planets from Perth which he did with excellence.

*In the ESA system then, the first step in mission selection was called a Phase A study. When was this carried out.*

There were two Phase A studies. The final one was completed in March 1978.

*Who were the ESA people that you came across during the pre-selection period, i.e. in the late 1970s?*

Roger Pacault chaired our meetings and sometimes Henk Olthoff.

*There was a lengthy political debate conducted in the scientific advisory committees of ESA during the late 1970s about the relative merits of Hipparcos, and its main competitor at that time, Giotto - a mission to*

*Comet Halley. There was a strong division at this time, almost along national lines: Germany was favouring Giotto, France favouring Hipparcos. How did you see this competition?*

There was a tight competition around Hipparcos. I doubted much that Hipparcos could win. I once said to a prominent astrophysicist, that there was only this one chance to get space astrometry. If Hipparcos was not selected this time there would not be a second opportunity. I personally would leave the project. He gave me the wise advice, not to use this as an argument, but only to use scientific arguments, and I followed his advice.

*So around 1980, how was Hipparcos perceived within the wider scientific community at this time? Were other astronomers generally interested, supportive, or excited?*

Astrometrists were generally supportive and in effect many worked later on the realization of Hipparcos.

*But it was still an uphill battle in getting it accepted?*

Oh yes, it was very uphill [8]. Only few astrophysicists declared their support and more were against Hipparcos. This is best illustrated by the crucial role of Edward van den Heuvel in the final decision of ESA's Astronomy Working Group in January 1980. He was a famous X-ray astrophysicist but he was a strong advocate of Hipparcos for its scientific value, when he was asked to present Hipparcos in the group. Before the meeting he had been able to convince two of the astrophysicists to vote for Hipparcos which then got the majority. That was a great surprise.

Without Ed van den Heuvel, Hipparcos would have lost to the EXUV mission (EXtreme UltraViolet) and nothing could have changed that decision. I have written a report [8] with documentation of these events. I did not hear about these events until much later when I met van den Heuvel and he told me.

*Around the 1980s, astrometry played an important part in American science, in part because of its relevance to solar system*

*navigation. One of the big players in US astrometry at the time was the US Naval Observatory based in Washington. How did they react to Hipparcos?*

Yes, they were interested and I always gave a talk there when I visited the US.

*In your own mind, did you see Europe as being in competition with the US at that time?*

No, there was no competition from US because astrophysics was so dominant there.

*You've said in your more recent writings that a mission like Hipparcos would never have been undertaken in the US. What did you have in mind when you said that?*

Could NASA have realised a Hipparcos-like mission? NO! For two reasons: the American astrometric community had much less resources of competence to draw from than were available in Europe, and secondly, as an American colleague said to me in those years: "You can convince a US Congressman that it is important to find life on other planets, but not that it is important to measure a hundred thousand stars."

*Which other nations were interested or supportive of space astrometry at that time - UK, Russia, other Scandinavian countries?*

There was much interest for Hipparcos in Russia, or USSR as it was. For example they published the first translation of an article on Hipparcos by me.

*You have also said in your own notes and memoirs that if Europe had not accepted Hipparcos in 1981, it would have been a lost opportunity that might never have re-appeared. What did you mean by that?*

It appears that the approval by ESA could well have failed, in which case I am sure Hipparcos would never have been realised. This proposition has been countered by a colleague: "You can never know that, something could have happened." But please consider the situation of astrometry at that time [10]. For decades up to 1980 the astrometry community was becoming ever weaker, the older generation retired and very few young scientists entered the field. I myself would have lost the faith that the astrophysicists would ever let such a space mission through, and others would also have left the field of space astrometry.

*Do you think that a revival later would have been possible ?*

If someone would have tried a revival of the idea one or two decades later, the available astrometric competence would have been weaker, and where should the faith in space astrometry have come from? When Hipparcos became a European project in 1975 and the hopes were high for a realization, the competence from many European countries gathered and eventually was able to carry the mission. This could not have been repeated after a rejection of the mission.

*Very interesting!*

**Politics and influence**

tm=43:37

*You have received some significant international recognition for your work over the years. In 1999, you were awarded the European Space Agency's Director of Science Medal for your leading contributions to Hipparcos. In 2005, the International Astronomical Union named an asteroid after you: Erikhoeg. In 2009, you received the Russian Pulkovo Observatory's Struve Medal for your contributions to astrometry. In 2014, you received an Honorary Doctorate from the Ukrainian National Academy of Sciences in Kiev. And in 2019, you received the Instrumentation Prize of the German Astronomical Society, for your lifelong work in astrometry. How have your contributions been viewed in your home country of Denmark?*

In Denmark there was strong support for astrometry and that put me on the path in science. Danish funds supported my work on space astrometry, but all resources for astrometry of the Copenhagen Observatory went into the automatic meridian circle in Brorfelde which was later moved to La Palma. That was OK and a great success in its own, but with the result that the observatory had no resources for space astrometry, except that I was of course allowed to do this work.

*Was Tycho Brahe someone whose achievements inspired you in any serious way? I recall that we had a wonderful early meeting of your data analysis team on Tycho's island of Hven.*

I have often read about Tycho Brahe and a few years ago I wrote a paper [12 and 13] about his fruitful interaction with the Landgrave in Kassel who was a bit older than he. It was surprising for me to see that the observations in Kassel were more accurate than those on Hven, but they were of fewer stars and they were published many years later too late to have any impact on the progress of astronomy.

tm=45:55

*Can you point to some specific strength or personal quality that's helped you on your scientific journey?*

I think of myself as a kind person interested also in the well-being of others. Sometimes I was too kind and too patient and then realized I was being misused. But sometimes my patience was so stressed that I lost my temper which I regret.

*That final comment surprises me. I have known you for forty years and worked very closely with you for more than thirty and I do not think I ever recall you losing your temper. You always spoke with passion with honesty and I think as a scientist one cannot do more than that!*

*Did you look back to anyone in particular who inspired you, encouraged you, or influenced you early on in your career?*

I would mention my teachers in Copenhagen as a student: Peter Naur and Bengt Strömgren as I have already explained.

*Is this something that you think about today - inspiring or encouraging young people setting out on their own paths in science ?*

Look around and be kind to yourself and to others.

**Closing off**

*This has been a fascinating look at just a few aspects of late 19th century astrometry pre-Hipparcos. And I hope we can talk about your later work on Hipparcos and Gaia in later interviews. Meanwhile, Erik Høg: thank you very much for joining me today.*

It has been nice for me to think again through all these eventful years as led by your questions Michael. I enjoyed very much the good atmosphere in the teams. It is great luck to look back on a good life and still be active in my family and in astronomy and also to be at good health.

tm=48:00

*Thank you again Erik.*

## ***References***

References are given about the history of astronomy related to space astrometry in the early 1970s and about Erik Høg. Most papers by Høg may be found on ADS by searching for "Høg, E." or "Hoeg, E.".

[1] Høg, Erik 2017, Young astronomer in Denmark 1946 to 1958.
In: Wolfschmidt, Gudrun (Hg.): Astronomie im Ostseeraum – Astronomy in the Baltic. Proceedings der Tagung des Arbeitskreises Astronomiegeschichte in der Astronomischen Gesellschaft in Kiel 2015. Hamburg: tredition (Nuncius Hamburgensis – Beiträge zur Geschichte der Naturwissenschaften; Band 38) 2018 (30 pp). http://arxiv.org/abs/1512.01924

[2] Høg E. 2014, Astrometry 1960-80: from Hamburg to Hipparcos.
Proceedings of conference held in Hamburg in 2012, Nuncius Hamburgensis, Beiträge zur Geschichte der Naturwissenschaften, Band 24, 2014. http://arxiv.org/abs/1408.2407

[3] Høg E. 2017, GIER: A Danish computer from 1961 with a role in the modern revolution of astronomy.
In: Nuncius Hamburgensis, Volume 21, 294-321. Editor Gudrun Wolfschmidt. https://arxiv.org/abs/1704.05828

[4] Kovalevsky J. 2009, The ideas that led ESRO to consider a space astrometry mission.
https://www.astro.ku.dk/~erik/Kovalevsky2009.pdf

[5] Høg E. 2008, Lennart Lindegren's first years with Hipparcos.
http://www.astro.ku.dk/~erik/Lindegren.pdf

Here follow three reports on the early history of Hipparcos from 1964 to 1980:
[6] Høg E.2017, Interviews about the creation of Hipparcos. 20 pp including an appendix of 8 pp.
http://www.astro.ku.dk/~erik/xx/HipCreation.pdf and it is contained in [9]

[7] Høg E. 2018, From TYCHO to Hipparcos 1975 to 1979. 21 pp.
http://www.astro.ku.dk/~erik/xx/Hip1975.pdf and it is contained in [9]

[8] Høg E. 2017, Miraculous 1980 for Hipparcos. 9 pp.
http://www.astro.ku.dk/~erik/xx/HipApproval5.pdf and it is contained in [9]

The three reports [6], [7], [8] are also placed in one submission to arxiv:
[9] Høg E. 2018, Astrometry history: Hipparcos from 1964 to 1980. 52 pp.
http://arxiv.org/abs/1804.10881
http://www.astro.ku.dk/~erik/xx/Hip64arXiv.pdf

[10] Høg E. 2011, Astrometry lost and regained.
Baltic Astronomy, Vol. 20, 221-230, 2011.
http://esoads.eso.org/abs/2011BaltA..20..221H and at:
http://www.astro.ku.dk/~erik/xx/2011BaltA.pdf

[11] Høg E. 2015, The Baltic Meetings 1957 to 1967.
In: Nuncius Hamburgensis, Volume 38 (2018), Editor Gudrun Wolfschmidt. 11 pp.
http://arxiv.org/abs/1512.01925

[12] Høg E. 2016, The Landgrave in Kassel and Tycho Brahe on Hven. 2 pp.
In the proceedings of a meeting in September 2016 in Bogota: Astronomiía Dinamica en Latinoamerica in RevMexAA(SC) Vol 50, 2018
http://www.astro.ku.dk/~erik/xx/Erik2.Hoeg.Oral.Tycho.pdf

[13] Høg E. 2016, From the Landgrave in Kassel to Isaac Newton.
In the proceedings of a meeting in September 2016 in Bogota: Astronomiía Dinamica en Latinoamerica in RevMexAA(SC) Vol 50, 2018.
http://www.astro.ku.dk/~erik/xx/Erik3.Hoeg.Poster.pdf

2021.10.31:

# An interview about the revival of astrometry after 1980

*by Michael Perryman and Erik Høg*

ABSTRACT: Michael Perryman invited Erik Høg to this interview in October 2021. - The first space astrometry satellite Hipparcos was approved by ESA in 1980 and launched to a three year mission in 1989. This interview is about Hipparcos, and how it was developed to measure 100 000 stars more accurately than could be done from the ground, measuring positions, motions and distances. Erik Høg was one of the scientific leaders of Hipparcos in the 1980s and 1990s, active in the design and data analysis. He introduced an add-on to the satellite in 1981 called Tycho which made the mission even more successful. This resulted in the Tycho-2 Catalogue in 2000 of the 2.5 million brightest stars in the sky with accurate positions and motions and including two-colour photometry. Hipparcos was followed by the launch in 2013 of its follower Gaia which is presently revolutionizing astronomy. This report contains the questions by the interviewer, the answers by Erik Høg, a link to the audio recording and references.

---

Here follows a script of the recording. Erik in red is quoted closely but Michael not so closely, with slanted text.
The audio is here: http://www.astro.ku.dk/~erik/xx/hoeg2-final.mp3 ,
but you should rather open it on Michael's website with photos and other interviews: https://www.michaelperryman.co.uk/gaia-project-interviews
The minutes and seconds are tagged at tm=... Total duration 45 minutes.

**Introduction**

Michael Perryman:

*Gaia is a scientific satellite of the European Space Agency launched in 2013. Today it is still thoroughly scanning the dark sky from a deep space orbit more than a million km from Earth. It is in the process of constructing a remarkable three-dimensional survey of more than two billion stars throughout our galaxy. I'm*

*Michael Perryman, and in a series of conversations I am talking to scientists involved in different aspects of the mission.*

*Joining me today - and for a second time - is: Erik Høg, Professor Emeritus of the Niels Bohr Institute, University of Copenhagen.*

*Erik was one of the people responsible for getting the Hipparcos astrometry mission adopted in the 1970s he was one of the scientific leaders of Hipparcos in the 1980s and 1990s and he was central in the developments of Gaia in the 1990s and 2000s.*

*In an earlier conversation we talked about the period leading up to the adoption of Hipparcos in 1980. Today, we're talking about the Hipparcos mission itself - from its adoption in 1980 to its completion in 1997.*

*What was the role of scientists in the overall development of ESA's Hipparcos mission?*

*What was involved in the preparation for - and the analysis of - the data sent down from the satellite? And what were some of the big ideas, and biggest challenges, involved in carrying out this pioneering space mission?*

*Let's find out!*

*Erik Høg: a very warm welcome again.*

Thank you Michael, I look much forward to speak with you again about these great times.

**The early steps**

*After several years of scientific preparation, which we discussed in our first conversation, the Hipparcos space astrometry mission was eventually accepted by ESA's advisory committees and confirmed by ESA's Science Programme Committee in Paris in September 1980.*

*Do you remember where you were when you heard this news, and how you felt?*

No I do not remember where I heard this great news, perhaps because in most of September that year I was in China, invited by Mrs Yeh Shi Hua and Dr. Hu Ningsheng. It was the first of many visits and China became part of my life, but that is another story. Hipparcos stayed as my top priority for over twenty years.

*But how did you react to the news when you heard of its acceptance - excitement, relief, or what?!*

I do not exactly remember, but I am sure I was happy - and perhaps a bit surprised as I have already explained.

Yes because it was a long struggle to get the mission accepted.

*Before we talk more about the work of you and your scientific teams, let me summarise briefly for our listeners, how these sorts of space science missions are organised.*

*ESA appoints a Project Manager (in this case Franco Emiliani) responsible for the technical development and a Project Scientist responsible for the scientific aspects, and ensuring that the science goals are achieved obviously these systems have to work closely together, respecting each other's constraints and goals.*

*ESA eventually selects an Industrial Consortium responsible for building the satellite. And as Project Scientist for Hipparcos, I set up a scientific advisory team to guide all aspects from start to finish. At the same time, ESA approved the setting up of three scientific consortia across Europe, one to compile the catalogue that would be the starting point of the satellite observations and two independent teams to process the satellite data and create the final catalogue.*

*So, in 1981, we were sitting with a pile of technical notes which resulted from several years of theoretical studies about the mission's feasibility. What do you recall of the first steps that ESA took in moving from these studies to a real mission?*

What first comes to my mind is the invitation to a meeting in 1981 in Paris with engineers from MATRA, one of the a companies competing to build the satellite. They wanted to gather an advisory team of scientists, which they did. We often met in Toulouse over the next few years. This was under a contract where we promised confidence because there was a competing company also hoping to build the satellite. ESA did not much like such advisory teams, but they were tolerated.

Back to the first meeting with MATRA where I was the only scientist yet. They impressed me by showing a number of designs of the optical telescope system. Three of the designs were definitely better than the one we had considered in our Phase A studies: I now understood how clever industry can be in optical design.

And in the end MATRA obtained the contract for Hipparcos.

*Yes, and we benefitted from their expertise and experience in many other areas as the project progressed…*

**The Hipparcos Science Team and pre-launch**

*From 1981 to the end of mission in 1997, the Science Team numbered about 12 people from across Europe covering different scientific, astronomical, and technical competences we met in person every 4-6 months, reviewing progress, and assigning tasks and priorities. You were one of ESA's (my) scientific advisers throughout this time and you also led one of the two data processing teams, which you called NDAC.*

*Let's talk first about the Hipparcos Science Team - which I chaired for 17 years. Tell us about these meetings from your perspective. What sort of topics were handled by the Science Team and what sort of preparation did this demand from your side?*

I remember that a great deal of my time those years was dedicated to this work. My wife believes it was never out of my mind. We never took more than one or two weeks summer vacation. Not until 1992 did I take as much as three weeks, enough to bring us by train and ship as far north as Tromsø at the polar circle in Norway.

In preparation of the meetings I read a lot of documents on technical or organisational matters, and we prepared documents for the Science Team to promote our ideas, and to present the mission at various meetings. I enjoyed the meetings and I was always fully alert, while I sometimes saw a dreaming engineer when science was discussed.

I regret that I was unable to take good notes while a few of my colleagues wrote with a perfect hand in solidly bound books. I took notes but only on a piece of paper. I have sometimes asked Lennart Lindegren what happened at a certain meeting and he just looked it up in his book. The chairman, you Michael, always had a carefully planned agenda which was followed, and tables were always booked in a nice restaurant for the evening. I enjoyed these social occasions very much. I should add that we all shared the bill for the evening, but ESA paid our travel expenses.

Our Science Team meetings sometimes took place at the home institute of one of the team members. Such a meeting in Copenhagen had a session at the location where Ole Rømer put up the meridian circle he had constructed in 1704. This place was 20 km outside Copenhagen because Rømers official observatory on top of the famous Round Tower was too unstable for his delicate observations of positions.

tm=9:00

*Most of our meetings were at ESA's Research & Technology Centre at Noordwijk, in The Netherlands. There was always a lot of technical information provided by the project team and guidance requested from our scientific advisors…*

*Did you see a big change in the speed of progress starting in the early 1980s?*

I do not remember anything else than a nice speed of progress all the time after 1981.

*There is a big gulf between early ideas for a space science mission and the detailed design, the technical development, the integration and testing and of course the launch and operation. After the Phase A studies in 1980, which led to a baseline mission concept were there any major new technical developments made to the basic design?*

The baseline was maintained, but it was now implemented with reality, e.g. with a good optical design as I mentioned. Gas jets were introduced for attitude control instead of the reaction wheels that I had imagined.

*During the Phase A Studies you already made a number of innovative contributions to the instrument and satellite design. But in the period we are talking of now were there aspects of the instrument, or the satellite technology, that concerned you most, or where you felt you had the most influence in the years up to launch?*

I only remember that I always felt I was doing useful work on a great astronomical project.

*We want to focus here on some of the satellite data analysis aspects. Before launch, from 1981-1989, your consortium NDAC (for Northern Data Analysis Consortium) was focused on preparing for the satellite data…. first, why was it called NDAC?*

There were two data analysis teams who developed the software independently, and would process the satellite data independently - the other was the FAST Consortium, led by Jean Kovalevsky. We called ours the northern consortium because the FAST consortium was more in the southern part of Europe.

*What did these preparations for the data processing software involve, and how was your team actually structured?*

From 1981 we had to build a consortium as required by ESA. It was clear from the beginning that it should involve the Royal Greenwich Observatory where I had long had contact with Andrew Murray - whom I had brought into ESA's Astronomy Working Group (AWG) to succeed me. Also Mike Cruise from Mullard Space Science Laboratory was supporting the project. I had always seen Murray as the natural consortium leader, but he insisted that it should be me when we began forming the consortium called NDAC. Of course Lund Observatory with Lennart Lindegren should come in. In Copenhagen I had support from the Danish Space Research Institute, and from the Danish Geodetic Institute. In both cases we could get access to computing facilities and to methods of treating large systems of equations, especially through Knud Poder from the Geodetic Institute.

*When you were building up the data processing software for the treatment of the real data there were many studies and simulations that were undertaken and the results of these simulations and tests was also fed back to the design of the satellite…*

*What sort of results from these early studies was fed back into the satellite design?*

We undertook the first simulations in Copenhagen in 1980 which were crucial for the design of angular one-dimensional observations with an astrometry satellite with revolving scanning. We could show that the subsequent reconstitution of the celestial sphere could give (1) relative positions and (2) absolute trigonometric parallaxes, for a network of stars covering the entire sky, after only one year of observation. Many other simulations were made to develop the data reduction chain of NDAC.

*Yes, and these studies led to the choice of basic angle, piloting of the detector, gas jet firings for attitude, etc.*

*Let's mention the main task leaders of your consortium NDAC: these people stayed with the project for its entire duration!*

Many people were involved in the consortium: all 19 are listed by name in the final catalogue. Let me mention here those working on the data reduction during most of their time: Carsten Skovmand Petersen and myself from Denmark, Lennart Lindegren (Consortia Leader from 1990) and Staffan Söderhjelm from Sweden, with Dafydd Wyn Evans, Floor van Leeuwen and Andrew Murray from the Royal Greenwich Observatory, United Kingdom.

I always aimed at a collaboration of very few institutes: at RGO, Copenhagen and Lund for the sake of efficiency and ease of organisation, but efficiency was also a hard necessity with the very limited resources I could expect to have.

*Your approach to the data analysis problem was quite different to the structure of the other data analysis team, called FAST?*

Two consortia were working in parallel on the Hipparcos observations in order to compare the results from two independent groups for verification of these unique space observations. Jean Kovalevsky led the other consortium (FAST) and he created a much bigger organisation than I did for NDAC. He wanted to involve as many countries in Europe as possible to make Hipparcos a real European mission, a very laudable goal indeed, and he succeeded with a consortium of 82 persons in five countries: France, Germany, Italy, The Netherlands and (a few from) the US. In this way, Kovalevsky paved the way for further European space astrometry after Hipparcos.

*The limitations that existed in computers and computer networks at this time meant that the successive steps of the data analysis were made by physically* *sending* *the data from one institute to the next, tell us about this.*

In NDAC the satellite raw data were first treated at the Royal Greenwich Observatory (UK), and the output was sent on magnetic tapes to Copenhagen, where the abscissae of the stars along great circles were derived. They were used to derive the five astrometric parameters for all stars. At Lund, in Sweden, the processes were defined mathematically, and the double stars treatment also took place there, led by Staffan Sõderhjelm and Lennart Lindegren.

tm=17:00

**Tycho**

*When Hipparcos was accepted by ESA in 1980, it targeted measuring 100,000 stars to a typical accuracy of 2 milliarcsec, 2 thousandths of an arcsec we ended up doing a bit better than that in practice: almost 120,000 stars to an accuracy of 1 milliarcsec.*

*But a big development came in 1981, with your ideas for an add-on to the satellite, which you called Tycho. Let me read from my own account of the mission which gives the background:*

*In April 1981, already well into the satellite design phase when significant modifications would normally have been rejected out of hand for reasons of risk, and for the increased cost that they could incur, Erik Høg pointed out that a rich seam of satellite data was lying untapped. He realised that the signals from the satellite attitude detectors contained a staggering quantity of star positions that were, quite simply, not being sent to the ground. But once pointed out the*

*harvest was obvious: positions for the million or more stars not being observed by its main detectors. Working quickly to optimise the system, and with a little urgent lobbying, the small amount of additional funding needed was made available. A filter and additional detectors were added, an extra telemetry channel set up to send the data to ground, and a new data processing team put in place to handle the data stream.*

*So Erik, in addition to the two existing processing teams you set up another - the Tycho Data Analysis Consortium which you then led through to completion. How did this idea of using the satellite's star mappers to get this extra star catalogue come to you?*

The idea came in March 1981 when I wanted to define suitable meridian circle observations of reference stars for the Hipparcos attitude determination. I realized that the photon counts from the star mappers could be used to detect many more stars than the few thousands required for the attitude. My estimate was that positions and magnitudes could be obtained for at least 400 000 stars, and perhaps many more. But this required that all the data had to be transmitted to ground, about 100 Gbytes, the same amount as for the main Hipparcos mission. I immediately wrote three short notes to ESA about this Tycho project as I called it.

*This was really a very interesting example of something which was quite obvious to everyone as soon as you pointed it out, but had escaped everyones attention except your own. But I think that everyone on the Science Team immediately understood its importance?*

The proposal was well received by you Michael, and in the Science Team. I remember especially seeing the faces of Michael Grewing and Jean Kovalevsky at the meeting. The name Tycho was accepted without discussion, and this was a great relief for me since I remembered that Kovalevsky had introduced the name Hipparcos for the satellite instead of the name Tycho that I had wanted.

*Of course, convincing ESA to find and approve the funding for the changes was another matter?!*

Later in 1981, ESA approved the implementation and the extra cost as just mentioned by you, Michael. *From Vol.4 XI I read:* "Had the Tycho proposal come a couple of months later, the satellite design would have been frozen, and the idea of the Tycho project would have been a lost opportunity. This wonderful idea would have been difficult to forget about, even though we were immersed in the fascination of the main mission, and in all the work it gave us."

*The data reduction for the Tycho experiment still had to be set up and organised. So late on in the mission, and with a such a big data volume, that was not an easy challenge, Erik?*

No, that was not easy at all. I recall a moment at the beginning when I was overwhelmed having to do that and said so to Lennart Lindegren, when we were waiting for the bus somewhere. He just smiled and said he was sure we could do it, I then even believed he would do it in Lund!  Lennart was always a great support. I never hesitated Michael, when I had the good ideas in those years, and luckily I did not know how big the tasks would be.

Michael Grewing was the first to tell his interest by letter when I proposed the Tycho project, and as director of the astronomical institute in Tübingen he maintained an excellent young team there over the years. Carlos Jaschek was director of the institute in Strasbourg, and a team there also did a wonderful job. But these two teams, together with Copenhagen and Lund, were still not enough for the task ahead.

Derek McNally in the UK made an effort to form a team, but continued support for a team failed. So we had a big problem. But then Roland Wielen as director of the Astronomisches Rechen-Institute in Heidelberg (who had recently taken over from Walter Fricke) he offered to establish a team and of course I gladly accepted.

Uli Bastian of that team told me later that at first he did not want to work on Tycho because he was always an astrophysicist. But Wielen insisted: then he had to do it - and he did it wholeheartedly throughout the years.

*All this led to the Tycho Catalog of a million stars, published at the same time as the main catalogue in 1997. Tell us about the scientific team that you organized to process this additional data.*

The observations consisted of photon counts from the satellite's star mapper slits, continuously obtained during the 37 months of mission. Detection of a star in this data stream - or rather of a signal above the noise which could therefore be a star crossing the slits - that was recorded as a function of time. Stars were later confirmed and identified with real stars in the sky using two sources of information: 1) accurate knowledge of the pointing of the satellite telescope for every moment of time, the so-called satellite attitude, and 2) a list of 3 million stars with positions known from ground based observations.

This may seem fairly simple in principle, but it was quite complicated in practice, because the work had to be done within a tight time schedule and it had to be divided between three institutes due to the limited computer capacity and man-power. It is fair to say that no astrometric research before had been done under such constraints as the Hipparcos and Tycho data reductions. Therefore, we owe a cordial thank you to all those working in the consortia for their dedication to our great projects during many years.

*In addition to new people that you brought in to the Tycho Consortium, many individuals already working in the NDAC and FAST teams also contributed expertise and data?*

Yes, indeed. The Tycho Input Catalogue of 3 million stars was based on early access to the Guide Star Catalog for the Hubble Space Telescope as provided by Brian McLean and Jane Russell. The satellite attitude was provided - from the main satellite data processing - by Franceso Donati and Pier Luigi Bernacca in Italy (from the FAST Consortium) and by Floor van Leeuwen in the UK (from the NDAC Consortium).

Many people were involved in the Tycho consortium: all 37 are listed by name in Vol. 1 of the final catalogue. Let me mention those working on the data reduction during most of the time: Claus Fabricius, myself, Valeri Makarov, Holger Pedersen, and Carsten Skovmand Petersen from Denmark, Daniel Egret, Jean Louis Halbwachs, Jean Kovalevsky, Catherine Turon and Pierre Didelon from France. Further members were: ten from Germany, two from Italy, two from The Netherlands, one from Sweden, one from Switzerland, four from United Kingdom, and two from the USA.

*Among the many Tycho Consortium meetings that you held over the years, do any stand out in particularly in your memory?*

We had many meetings of the consortium during the 1980s at the participating institutes, and one of the meetings took place in 1984 on the island Hven, the historic site near Copenhagen where Tycho Brahe with his many collaborators made his famous observations of one thousand stars during twenty years from 1576-96. At the meetings we discussed the progress of work and defined how to proceed - and we had a nice and productive time together. I stayed a year as guest professor at the institute in Tübingen and two months in Strasbourg.

tm=28:00

We mentioned that the Tycho Catalogue was published in 1997, at the same time as the main Hipparcos catalogue. But this was all followed by the improved Tycho-2 Catalogue in 2000. What was the difference between Tycho-1 and Tycho-2, and what was needed to construct it?

One big difference for our work on Tycho-2 beginning about 1997 was that we collected all star mapper detections from the whole mission close to any star in the input catalogue. From these fifty or more detections we could decide whether we had a real star with much better statistical certainty than from the single detections used for Tycho-1 and this resulted in accurate positions for 2.5 million stars, a fantastic number for that time.

This possibility had been clear to me all the time, but it had only become possible with the greater computing power we could now buy. We were also able to get the needed private funding for the computer and for Claus Fabricius and Valeri Makarov to do the work in Copenhagen. Andreas Wicenec supplied the satellite data from Tübingen on a new kind of tapes. Uli Bastian and Peter Schwekendiek in Heidelberg supported with computations, and the final satellite attitude was provided by Floor van Leeuwen at the Royal Greenwich Observatory.

And then, very importantly, there are good proper motions for every star in Tycho-2?

Yes! The other big advantage over Tycho-1 is that Tycho-2 indeed contains accurate proper motions. They were based on the Tycho-2 positions, which were then combined with a new analysis of the positions observed during the past 100 years and given in 144 ground-based astrometric catalogues provided by the US Naval Observatory in Washington DC by Sean Urban.

I recall meeting Sean at a conference in 1997 in Gotha and realizing this fantastic possibility. They had not yet completed the new analysis but we received all the positions just in time to derive the proper motions for our publication in 2000. Had they come later our funding would have ended before we could do so. Tycho-2 has been the most cited astrometric catalogue ever since 2000 apart from the Hipparcos Catalogue.

Close double and multiple stars could also be treated much better in the Tycho-2 processing. Claus Fabricius and Valeri Makarov published 2-colour photometry for almost 10 000 components in systems with separations above 0.3 arcsec.

tm=32:00

And the Tycho-2 Catalogue took on a renewed importance with the advent of Gaia, because the positions from Tycho-2 from the early 1990s could be combined with the early Gaia positions to give the TGAS (or Tycho-Gaia) Catalogue with almost 25 years of proper motions and which was also important for the early Gaia Astrometric Iterative Solution.

Did you have any involvement in this latest TGAS Catalogue?

No I did not, except sharing the joy over TGAS. The rate of citation for Tycho-2 has stayed constant even after publication of Gaia results beginning in 2017.

**Launch**

Hipparcos was launched by Ariane 4 from French Guyana in 1989. As one of the key scientists, and one of the consortium leaders, you attended the launch. Tell us about your experiences, seeing the launch, and witnessing this new chapter in astrometry .

The Hipparcos Science Team and many others witnessed the launch in Kourou, but some of us had the special honor to fly in the super-sonic Concorde. From Hipparcos we were the consortia leaders as well as Professor Pierre Lacroute, the father of space astrometry. We flew from Paris to Dakar and then on to Kourou, far above the highest clouds, and we could see the setting sun, all the time at the same angle above the horizon while we crossed the Atlantic. I had a window seat and noticed that the wall became warm during flight, due to air friction at the super-sonic speed.

We saw the impressive launch facilities in Kourou, had our luxurious breakfast at the beach, and we saw the launch from a spot very close to the rocket. We saw the fire and smoke and we saw the big rocket slowly lifting off. But we heard nothing, not until many seconds after. We were happy seeing the glowing fire from the rocket moving away up there, and we toasted in champagne.

After a spectacular and flawless launch, we had big problems - the apogee boost motor intended to circularize the satellite orbit failed to operate and the satellite was stuck in an elliptical 10-hour orbit rather than the planned geostationary.

I don't want to dwell on the enormous technical problems that followed from this but instead let's concentrate more on the treatment of the satellite data. Do you remember the moment when the first "reference great circles" were constructed and when the first sphere solutions or first Tycho results started to appear?

I do not recall precisely these great moments. But I remember a conference in 1989 of the Pulkovo Observatory near Leningrad (now St Petersburg) where I had been asked to present Hipparcos on behalf of the Science Team. It was during the months after the post-launch apogee boost motor failure, and everybody asked about the latest news from the great Hipparcos mission - which had a very uncertain future after the boost motor failure.

In the presentation I had shown a picture of the first Tycho observation. It was a plot with the very first recording of transits over the four slits in the star mapper during satellite commissioning. I had made a number of prints for distribution and I can still see all the eager hands reaching to get such a print of the first Tycho observation.

A distinguished person approached me during a break. Away from the crowd he presented himself as coming from the Russian space agency and offered to launch a second Hipparcos in case ESA could not.

And this is where you first met Valeri Makarov, who played a big part in the Tycho processing?

tm=36:00

Yes, Viktor K. Abalakin, director of the Pulkovo Observatory, introduced me to his young people: Mark Chubey and Valeri Makarov. I had before heard of Russian plans to launch a Hipparcos successor ten years later in order to derive very accurate proper motions from positions at the two epochs. Now I met these people with whom I had very fruitful discussions during the following years. Valeri Makarov soon became my very bright collaborator in Copenhagen as I've already mentioned.

A year later I was again invited to Pulkovo to tell about Hipparcos progress. They took me on a visit to Moscow to the famous space control centre, and further on to their observatory in the Caucasus. I enquired about their plan for a successor. After a day trying to understand their design, I realized that I could not understand how it would work, and that instead I was in the process of designing a better more accurate Hipparcos satellite. During the further travel I continued the design and discussed it with them. This led to further visits in Pulkovo, Moscow, Copenhagen and Lund for discussion of our ideas. Finally in 1992 I made a design with Lennart Lindegren of a mission with CCD detectors which we called Roemer and which later became Gaia [1].

I also recall a conference in Cambridge probably in 1990 when the first analysis of some Tycho observations had been made. I could then see that we would be able to make a catalogue of one million stars, not only 400 000. That was great news. The reasons were a slightly higher quantum efficiency of the photo-multipliers and a more accurate background determined with a method invented by Andreas Wicenec using the median value of the photon counts during an interval of time.

I think it's worth emphasising that the final catalogue accuracies, both for the main Hipparcos Catalogue, and for the Tycho Catalogue were very close to the theoretical predictions that you and your colleagues had made much earlier on in the mission design - for such a complex instrument, this was a great achievement?

I would rather say that the predicted performance was significantly exceeded in several respects. The final accuracy of Hipparcos reached 1 milli-arcsecond instead of the predicted 2 milli-arcsecond, and this is quite significant in the view of an astrometrist. And the Tycho-1 Catalogue contained one million stars - compared to the 400 000 in my original proposal. The Tycho-2 has even 2.5 million stars and also accurate proper motions. So Hipparcos performed much better than expected in spite of the bad orbit. We really must admire the responsible operations team for their technical expertise.

tm=40:00

**Finalising the catalogue**

The eight years from 1989, when the satellite was launched, to 1997 when the final results were published were very hectic, all of the teams working to a tight schedule, to get the results out to the wider astronomical community, and when research based on the data actually started.

You had handed over leadership of the NDAC Consortium to Lennart Lindegren soon after launch to focus your efforts on the Tycho data, and you presented the Tycho Catalogue for the first time, at a major international conference in Venice in 1997.

What was it like Erik to be in this position, and how did the wider astronomical community react to it?

The Venice conference was a great event in every respect, the setting itself in Italy where I knew that you and Pier Luigi Bernacca had done the perfect preparations. The results we could present were very well received.

I know you gave many lectures, and many interviews with the press, after the catalogue publication. Do any particular events stand out in your memory?

I think first of the big International Astronomical Union conference in 1997 in Kyoto, Japan, where we could present our results  - and we could again see they were very well received.

Looking back now Erik, In one lifetime you have helped take astronomy from the confines of the Earth out into space with two highly successful space missions.

You could never have imagined such progress when you started out in the field 60 years ago?

No, I never imagined anything like that. I began with astrometry at the new meridian circle in Denmark in 1953 when I was just 21 years old. In Copenhagen I saw that astrometry was central for astronomy. But soon after I understood that astrometry was not so highly rated everywhere. From 1958 at the Hamburg Observatory at first on a 10 months fellowship I wanted to do astrophysics because I clearly saw that real astronomy happens in that area while astrometry was rather old-fashioned.

But in July 1960 something happened. I had a revelation you might say, the vision of photon counting astrometry with meridian circles. It could be done using computers, punched tapes and all that. That brought me back to astrometry. It has turned out after all, that I have done more for astrophysics than I could ever have done as an astrophysicist.

The idea was well received in Hamburg-Bergedorf, the director Otto Heckmann saw that it should be used for the meridian circle expedition to Australia which was in preparation. And Heckmann trusted me, he trusted that I had the talent and that I had the stamina to hold on to the project for years. I had the luck to have the right idea at the right place and right time and I had the luck to meet the right people. The same luck I had later with the design of Hipparcos in 1975, the Tycho star mapper project, and the Roemer mission with CCDs which led to Gaia. It seems that the luck was there again with the proposal in 2013 of a Gaia successor in twenty years, a project now having a high ESA priority and a quite probable launch in 2045 [2].

As I mentioned in our first interview. You received some important international recognition for your work over the years:

In 1999, you were awarded ESA's Director of Science Medal for your leading contributions to Hipparcos.

In 2005, the International Astronomical Union named an asteroid after you: ErikHøg.

In 2009, you received the Russian Pulkovo Observatory's Struve Medal for your contributions to astrometry.

In 2014, you received an Honorary Doctorate from the Ukrainian National Academy of Sciences in Kiev.

And in 2019, you received the Instrumentation Prize of the German Astronomical Society, for your lifelong work in astrometry.

You have played a leading role in the move of astrometry to space, and inspired many others, but there is still far to go for the generations that follow?

Yes there is far to go and the people are there and generations will come, all well prepared.

**Closing off**

This has only touched on a few aspects of the Hipparcos mission in particular some of the main points seen through the eyes of one of space astrometry's leading scientists and it's been interesting to hear some of your thoughts as you look back over this important period.

Erik Høg: thank you very much for joining me again today.

Thank you Michael for inviting me to go through all this and for your perfect guidance. Thank you.

tm=45:31

## References

[1] Høg E. 2011, "**Astrometry history: Roemer and Gaia**"
26pp at: http://arxiv.org/abs/1105.0879

[2] Høg E. 2021, **The Gaia Successor in 2021.**
3 pp at: http://arxiv.org/abs/2107.07177 [astro-ph.IM]
http://www.astro.ku.dk/~erik/xx/GaiaSuccessor2021.pdf

Further references about the history of astronomy related to space astrometry in the early 1970s and about Erik Høg are given at the first interview. Most papers by Høg may be found on ADS by searching for "Høg, E." or "Hoeg, E.".

2022.02.19

# The billion-star astrometry after 1990

*by Michael Perryman and Erik Høg*

ABSTRACT: Michael Perryman invited Erik Høg to this interview in February 2022. - The second space astrometry satellite Gaia was approved by ESA in 2000 and launched to a five year mission in 2013. This interview is about Gaia, and how it was developed to measure millions of stars much more accurately than had been done with the first astrometry satellite Hipparcos, measuring positions, motions and distances. Gaia is presently revolutionizing astronomy. Erik Høg was one of the scientific leaders of Hipparcos and Gaia, active in the design and data analysis. His work with astrometry began in 1953, 69 years ago, at a meridian circle. In 2013 he proposed a successor to Gaia which has become a candidate for launch by ESA about 2045. The report contains the questions by the interviewer, the answers by Erik Høg, a link to the audio recording, and references.

---

Here follows a script of the recording. Erik in red is quoted closely but Michael not so closely, with slanted text.
The audio is available here: http://www.astro.ku.dk/~erik/xx/hoeg3-final.mp3 ,
but you should rather open it on Michael's website with photos and other interviews: https://www.michaelperryman.co.uk/gaia-project-interviews
The minutes and seconds are tagged at tm=.... Total duration 52 minutes.

## Introduction

Michael Perryman:

*Gaia is a scientific satellite of the European Space Agency launched in 2013. Today it is still thoroughly scanning the dark sky from a deep space orbit more than a million km from Earth. It is in the process of constructing a remarkable three-dimensional survey of more than two billion stars throughout our galaxy. I'm Michael Perryman, and in a series of conversations I am talking to scientists involved in different aspects of the mission.*

*Joining me today - and for a third time - is: Erik Høg, Professor Emeritus of the Niels Bohr Institute, University of Copenhagen.*

*Erik was one of the people responsible for getting the Hipparcos astrometry mission adopted in the 1970s he was one of the scientific leaders of Hipparcos in the 1980s and 1990s and he was central in the developments of Gaia in the 1990s and 2000s.*

*His career in astrometry has extended over a remarkable 70 years!*

*In our first interview, we talked about the period leading to the adoption of Hipparcos by ESA in 1980. In the second, we focused on Hipparcos itself, from its acceptance in 1980 to its completion in 1997. Today, we're talking about his work on Gaia, and on another new astrometry mission for the future.*

*So… What were the steps taken in advancing space astrometry after Hipparcos ? What were the challenges in optimizing some of the details of the Gaia payload? And what sort of future does space astrometry have in the coming decades?*

*Let's find out!*

*Erik Høg: a very warm welcome again.*

Thank you Michael for inviting me and all the others to these interviews on the different aspects of space astrometry, and also for writing the fine essays and placing all this on your website.

*Ah! You've found them interesting?*

Oh yes! For example, I was fascinated the other day listening to the interview with Bruno Sicardy about solar system occultations. He told how amateur astronomers with their 20-40 cm telescopes and CCD cameras observe light curves when a star is occulted by a solar system body. From these light curves Sicardy and his group in Paris have studied the atmospheres of Pluto, of asteroids or satellites of Saturn, and they have even discovered a ring system around an asteroid. This has become possible with the very accurate positions of stars from Gaia because the event can now be predicted to within seconds and also precisely where on Earth it will happen. This has activated many amateurs in Europe and around the world. It is just one of the many remarkable results coming from Gaia, and it reminds me of my own visual photometric observations in 1949 as a school boy of an eruption of the variable star SS Cygni with my homemade (12 cm mirror) telescope - as many other amateurs did in those years.

*That is a wonderful connection between your early days and the present results from Gaia...*

tm=3:40

## Early steps

*So, Gaia…*

*… a mission which is today revolutionising almost all areas of astronomy*

*… and which is, now, generating more scientific papers each year than any other ESA mission. See [4]*

*Where shall we start ?*

*I think we should go back to the 1990s,*

*and ask you to describe the interest in new missions that Hipparcos itself had generated?*

Yes, there was excitement and interest around the world to extend space astrometry to more stars and better accuracy. And there were many ideas.

*Can you recall some of these for us - there were a lot but it would good to show how just much interest there was at the time?*

Yes, in the US there were (POINTS, USNO - NEWCOMB, MAPS, FAME, AMEX, OBSS, JMAPS, OSI; by leading astrometrists, including Reasenberg, Duncombe, Jefferys, Seidelmann, Johnston, Hemenway +) and in USSR/Russia (Lomonossov, REGATTA-ASTRO, AIST, LIDA, OSIRIS; by Chubey, Makarov, Yershov, Nesterov **+**)

but even SIM (by Mike Shao) which did very well in the US Decadal Survey in 2000, was abandoned in 2010

DIVA (in Germany; by Siegfried Röser, Uli Bastian +) targeted 0.2 mas to 15 mag

JASMINE (by Gouda, Yamada +) is still being developed in Japan

in China, STEP (exoplanet detection) was proposed in 2013

and there were also other astrometric missions proposed to ESA: NEAT (M3) and Theia (M5).

A total of 19 names of ideas or projects are listed here.

tm=5:40

*Although we'll not go into details, all of these - except JASMINE - eventually fell by the wayside…*

*leaving Europe, again, in the "driving seat" in space astrometry.*

*Why do you think this was?*

There are several reasons. The ideas needed for space astrometry started in Europe already in 1960 with photon counting astrometry, and in 1975 the basic design of a scanning satellite was on the table - as we talked about in the first interview: It was photon counting astrometry in a telescope with a beam combiner imaging two fields of view scanning stars on great circles and covering the whole sky systematically. An input catalogue with 100 000 stars selected for the astrophysical interest should be observed during 3 years to obtain positions, proper motions and parallaxes with an accuracy of 2 milliarcsecond. That was on the table.

A year before that, in 1974 the early ideas for space astrometry had changed from being a purely French undertaking to become a European project within ESA. That opened a major opportunity for the still existing astrometric community in Europe to become active, and so they did.

tm=7:09

And things then moved quite quickly from these early ideas in Europe, to reality?

Yes, given the basic design and the active scientists, European industry was able to develop the Hipparcos satellite for approval in 1980 and launch in 1989. The European astronomical communities in ESA countries joined efforts to treat the observations and produce the final Hipparcos and Tycho Catalogues in 1997, and they surpassed all expectations with respect to accuracy and number of stars. All this gave a solid foundation in ESA, in European industry and in the scientific community for future space astrometry in Europe when new good ideas appeared.

It is characteristic that the development in Western Europe has followed a very straight path: From meridian circles to Hipparcos, then to Gaia, and now towards GaiaNIR which will probably be launched about 2045 as a large ESA mission.

tm=8:15

*Yes, it is an interesting point about the common development of the scientific and industrial aspects giving this enormous basis for future ideas. We talked about Hipparcos in previous conversations and we are going to talk today focusing on Gaia.*

*What were the first steps along the path to a new space mission in Europe after Hipparcos ?*

The first step towards a new ESA mission was taken in 1990. During a visit to the Pulkovo Observatory I discussed with my USSR colleagues their existing proposals for a successor to Hipparcos. This led me within a day to begin designing a better more accurate Hipparcos. This generated further visits to Pulkovo, Moscow, Copenhagen and Lund for discussion of the ideas with Lennart Lindegren, Mark Chubey, Valeri Makarov and Valodja Yershov. Finally in 1992, I made a design with Lennart Lindegren, from Lund, of a mission with CCD detectors which we called Roemer and which later developed into Gaia. Roemer was first presented at a Symposium of the International Astronomical Union in Shanghai in September 1992.

*It is worth noting that all this was going on fairly soon after Hipparcos had been launched, well before the final Hipparcos Catalogie had been published.*

*So what were the main features of this design, and how did it differ from Hipparcos?*

The proposal used direct imaging on CCDs with time-delayed integration, and it contained many other features implemented in the final Gaia. For a 5 year mission an astrometric accuracy of 0.1 milliarcsec was predicted at 12 mag, more than 10 times better than Hipparcos. The astrometric efficiency was 100 000 times higher, because CCDs have ten times higher quantum efficiency than photomultipliers, and because thousands of stars could be observed simultaneously, and this was obtained with the same telescope aperture of 29 cm as Hipparcos. Astrometry and multicolour photometry for 400 million stars were also included.

tm=10:50

*Going on in ESA at this time was the long-term planning for European space science, called Horizon 2000, which had been initiated by ESA's Director of Science, Roger-Maurice Bonnet.*

*You saw the next opportunity for space astrometry with ESA's call for a new mission in 1993?*

The Roemer proposal had aroused much interest in the Hipparcos Science Team. I should mention that the Hipparcos Science Team, which supervised all aspects of the Hipparcos mission until its catalogue publication in 1997, was still the main forum for top-level discussions amongst the European teams involved in space astrometry.

The ideas in Roemer were adopted in a mission proposal submitted to ESA on 24 May 1993 for the Third Medium Size ESA Mission (M3). The proposers from seven countries were: Kovalevsky, Lindegren, Halbwachs, Makarov, Høg, van Leeuwen, Knude, Bastian, Gilmore, Labeyrie, Pel, Schrijver, Stabell, and Thejl. We proposed to measure 100 million stars and to obtain an accuracy of 0.2 mas at V=13 in a 2.5-yr mission with a 34 cm telescopic aperture. At the faintest magnitudes 15-17, accuracies would be about 1 or 2 milliarcsec. So, a similar accuracy to Hipparcos but much fainter.

This mission was not one aimed at solving some particular problem, but rather to explore the rich complexity of Nature in order to find new patterns, anomalies and even contradictions to existing ideas. This was the spirit of the proposed ROEMER mission.

tm=12:50

*So it was very broad in its reach? But you did give some details about the scientific importance and objectives?*

Yes, the scientific objectives were detailed in ten pages of the proposal, covering stellar structure and evolution, stellar kinematics, cosmic distance scale, double stars and unseen companions, solar system objects, general relativity, and the celestial reference frame. The proposal, compiled and edited by Lennart Lindegren, had 28 pages. And there were two annexes, which I edited, running to more than 60 pages.

ROEMER was also an acronym for *"Rotating Optical Observatory for Extreme Measuring Efficiency and Rigour"*. A section called *"The FIZEAU option"* was included *"to point out a possible improvement towards a scanning satellite with ten times the angular accuracy of ROEMER"*. The section described a larger scanning satellite with *"two confocal Fizeau-type (or 'wide field') interferometers whose axes form a basic angle of the order 140 deg."*, that was a description fitting very well to the later GAIA. But the included optical system underwent major development before it was called GAIA, an acronym for Global Astrometric Interferometer for Astrophysics.

*The Roemer proposal was submitted to ESA in May 1993, as you said…*

*and then it was reviewed by ESA's Astronomy Working Group, the AWG…*

*What was the result of their deliberations?*

We heard to our delight that our proposal was rated the best scientifically among all astronomical proposals for M3. But the bad news was that it was considered to come too soon after Hipparcos and it was not sufficiently ambitious with respect to accuracy. It was therefore referred to a Cornerstone Mission study if 10 - 20 μarcsecond accuracy could be demonstrated.

tm=15:10

*So in 1993 Roemer really "lost out" to the cosmic microwave background mission Cobras/Samba later called Planck*

*What did you think of that?*

At that time I did not think of the competing and winning mission, only that we had lost.

*So the Astronomy Working Group considered Roemer's science very important for Europe,*

*but that the mission was not "ambitious enough"… Is that how we can summarise it?*

Yes, that is also what I remember. I thought it was a severe blow to a good proposal and that a ten times better accuracy was rather unrealistic. But you and Lennart had the courage to think bigger and your response built on ESA's request for interferometry.

*So what came next?*

## GAIA and Gaia

The major advance was the proposal led by Lennart Lindegren and yourself, Michael to use a small-baseline interferometer, housed within the Ariane 5 fairing, to achieve better accuracy and bigger mirrors to get to fainter magnitudes. Although the idea of an interferometer was later dropped due to technical complexity, the <u>advantage</u> was that interferometry had been foreseen within the Horizon 2000 cornerstone plans, and this really led to the mission gaining support; <u>both</u> because of its science goals, but also because of its technological appeal - even though other interferometry groups, especially infrared science - were still pushing their own ideas.

The reply to the ESA call for proposals of Cornerstone studies was submitted on 12 October 1993, a proposal to study for astrometry *“a large Roemer option and an interferometric option”*, GAIA. They should be studied as two concepts for an ESA Cornerstone Mission for astrometry *“without a priori excluding either”*, as Lindegren wrote in the cover letter. But in those years we soon considered interferometry to be very good for our purpose and the Roemer option was forgotten for several years.

*And you soon set out your own big ideas?*

In fact, I was also soon able also to "think big" and proposed in August 1994 a larger Roemer mission, called Roemer Plus, with larger apertures to obtain the high accuracy. It used direct imaging and no interferometry. But it attracted no interest because everybody, including myself, had come to believe that interferometry was the right way to go. But it was NOT at all, as industry showed us in January 1998.

In 1995 I proposed a new optical design GAIA95 with Claus Fabricius and Valeri Makarov, it was a Fizeau interferometer with Gregorian telescope, providing dispersed fringes for simultaneous astrometry and spectrophotometry. This design was immediately adopted for a minisatellite project DIVA by the German space agency DLR. The project was started and led by Uli Bastian and Siegfried Röser and was joined by US participation, but it was abandoned in 2002 after seven years when one of the German funding partners had dropped out. Luckily so we must say, because our German colleagues could then come back and join efforts to develop Gaia.

Ten years ago I have told the two fascinating stories of *"Roemer and Gaia"* and of *"Interferometry from Space: A Great Dream"* in two reports [1] and [2] available on arXiv.

*As well as your technical contributions to Gaia, which we’ll return to in just a moment,*

*you also looked at some specific <u>scientific</u> objectives*

*One was figuring out what magnitude Gaia had to reach to measure specific Galactic populations ?*

Yes, I prepared a table of *"Some Galactic kinematic tracers"* which was placed as the first table in the Gaia Study Report in 2000. It was elaborated during the years since the earliest discussions I had in 1994 at the IAU Assembly in the Hague. It contains selected tracers with the limiting magnitude and astrometric information necessary to study them, and it *"demonstrates that GAIA will provide adequate precision to meet the scientific goals"*. The selected tracers belonged to all Galactic populations: clusters, disk, halo and bulge, and the table was useful to justify the 18-20 mag completeness limit.

tm=20:40

*Another of your scientific ideas was your work on astrometric microlensing and what could be done with Gaia … can you describe this?*

The possibility to detect both photometric and astrometric effects of microlensing was discussed in the ROEMER proposal in 1993. The effect of relativistic gravitation appears when a massive foreground object, called a MACHO, passes in front of a star further away when this star is observed close to the crossing, and given the billions of photometric and astrometric observations by the proposed mission, the effect might actually be detectable. Such observations could potentially be used to determine the mass of the foreground object, the MACHO.

In 1994 I had a collaboration with the Russian cosmologists I.D. Novikov and A.G. Polnarev and I showed that all six physical parameters of a MACHO (mass, distance, two position coordinates and two proper motion components) could in principle be derived from observations of an encounter event with a star for which the position and proper motion are known.

This was a nice theoretical result, but the practice is difficult. However, in fact twenty years later, the detection of events with Gaia and photometric follow-up observations from a network of telescopes on the ground has become a rich source of science. Over a thousand transient sources had already been detected with Gaia data up to October 2016.

*It is a wonderful example of scientific discoveries coming out of this.*

tm=22:30

*Going back to the important challenge of optimising Gaia, can we talk about two of your main activities.*

*The first of these, was your work on how to best sample the pixels of the CCD focal plane, and you contributed many technical notes on this over several years?*

Yes, I dedicated much effort to this optimisation. The problem was that not all the billion CCD pixels could be sampled and sent to ground, because the read noise would be excessive and the telemetry rate was limited. So a clever selection of a subset of pixels was required onboard.

*And you showed that even at 20-21 mag, much of the sky is empty of stars, so its not useful to send either?*

Yes, much of the sky is empty, even at 20-21 mag, but there are other complications from diffraction spikes, and double and multiple stars, and the sky is not so empty inside clusters and galaxies. All this had to be optimised so that the best science could be extracted from the data in powerful computers on the ground by clever astronomers and technical experts. We also took care that the environment of all detected objects could be mapped in an area of about one square-arcsecond in order to detect companion stars or a nebulosity around the object which might itself be a compact galaxy or a quasar.

*You had some other people working closely with you on this?*

Yes, much of this work on sampling was done in collaboration with Frédéric Arenou (from Paris) and Jos de Bruijne (working under you in ESA), and the work was recorded in a series of technical reports. There were actually a total of 64 reports since 1997 with authors from Copenhagen about sampling, detection and imaging for Gaia. The final sampling used in the satellite operations built on our work, but was further optimised by industry and especially Jos de Bruijne at ESTEC.

tm=25:10

*Another of your major activities in optimising the Gaia payload was your work on the photometry.*

*Let me introduce this by saying that, from the earliest ideas for Gaia, I was very clear that measuring a billion or more objects was made far more powerful if we could also record their colours or spectra.*

*Tell us first about the scientific reasoning behind this?*

Sure, photometry was important, in fact for two different reasons. The instrument gives small but significant chromatic errors in the astrometry which can be corrected if a colour index of the star is known, and we need to know that for every transit for the sake of variable stars with changing colours. The other reason is astrophysical: the colour gives its position in the Hertzsprung-Russell diagram, and also (if the filters are carefully chosen) the spectral type, metallicity, and reddening. Many other photometric surveys are going on but we needed a reliable source giving prompt results: So it had to be by Gaia itself.

*Again, there were several key people helping you to think through the objectives, and the possible solutions?*

My colleague in Copenhagen, Jens Knude, was my constant support and he brought his lifelong dedication to photometry, especially with the Strömgren uvby system with intermediate width bands suited for early type stars.

People in Lithuania were also keen to support, in fact Vytas Straizys wrote to me already in 1994. He was director of the Vilnius Observatory and had developed the "Stromvil" system suited for classification of all spectral types, also in the presence of interstellar reddening. So the experts joined our efforts very early. For photometry and spectroscopy, the first idea was to add a third telescope with a collecting aperture of 0.75 x 0.70 $m^2$ in addition to the two large telescopes for astrometry. This was the design by Matra Marconi Space at the end of the industrial study in mid 1998.

*Various possibilities were considered, ranging from multiple filters, to low-resolution spectroscopy*

*How did you advance this particular study, and what was the final solution adopted?*

When the cornerstone study started, a Gaia Science Team was formed in 2001, under your leadership Michael, with various working groups devoted to studying specific aspects. One of these was the photometry working group with Carme Jordi from Barcelona and me as co-leaders. We had many special meetings on photometry and wrote many reports in those years. In the first design, photometry in four broad bands was obtained in the astrometric telescopes and in seven medium-width spectral bands in the smaller telescope. The next design, which was described in great detail in The Concept and Study Report to ESA in 2000, increased this to eleven intermediate bands. In 2006 our work ended with a paper of 25 pages in the journal Monthly Notices of the RAS (*not* Astronomy & Astrophysics as I said), with 35 authors. The number of bands had increased to a total of 19.

tm=29:30

*But that wasn't the end of this long and important story?*

NO! All this changed dramatically. In fact it changed while we were finalizing the MNRAS (*not* A&A) paper, but only a few of us knew that. The change had to be kept confidential because it was taking place during the competitive bid for the industrial contract for Gaia. One of the competitors, Matra Marconi Space (who eventually became the winning contractor), was proposing to simplify the design for reasons of mass and cost. They dropped the dedicated photometric telescope and reduced the number of focal planes from four to only one.

They asked you Michael whether it would be acceptable to replace photometric filters with two low-dispersion prisms. They reluctantly accepted that you engaged also me, and I also wanted to have Carme Jordi, Anthony Brown from Leiden, and Lennart Lindegren involved. It was a complicated problem, but eventually a very careful optimisation of the prism properties was done, resulting in the very satisfactory Gaia photometry that we have today.

*One of the points I wanted to get across in these conversations is that having an idea for a space mission is one thing…*

*but there is usually a difficult journey in convincing all the others that need to be convinced;*

*and a longer and even more arduous journey in designing and optimising an instrument*

*and especially a space instrument, where you have just the one chance to get it right!*

*We embarked on the ideas for a follow-on mission starting in the 1990s, in parallel with Hipparcos*

*and your ideas, and your commitment to the end goal, contributed so much to Gaia*

*But there comes a time when our main work is done, and we hand over to others?*

Yes, after leaving the Science Team in 2007, I was no longer involved in Gaia affairs. No Danish involvement in Gaia data reduction was conceivable. I was 75 enjoying my pension, and the institute was entirely focused on astrophysics, which I find very reasonable. Claus Fabricius and I are therefore the last Danish astrometrists in the row beginning with Tycho Brahe.

The Copenhagen University Observatory no longer exists. It started on top of the famous Round Tower in Copenhagen built in 1642 where Christen Sørensen Longomontanus, Tycho Brahe's assistant on Hven, had begun his work. And the Brorfelde Observatory, 50 km from Copenhagen, where I made the very first observations with the new meridian circle in 1953, was closed for science when the whole staff moved to Copenhagen in 1996. In 2005 the 363 year old Copenhagen University Observatory was absorbed in the Niels Bohr Institute - without any celebration or funeral.

*Well, your formal involvement ceased, and the hectic workload associated with it ceased…*

*but you've maintained a big interest in Gaia, and indeed in the role and future of space astrometry?*

I am doing nearly all my work from home, but I kept an office in the institute until 2019 and often visited the cosmology group for talks. I wrote a popular article in Danish together with two cosmologists, Peter Laursen and Johan Samsing, about the beginning of the Universe, and more recently an article focused on explaining the cosmic horizon. I am happy that Johan Fynbo and his student Kasper Heintz adopted and implemented my idea of detecting quasars by their zero proper motion using Gaia results.

tm=34:00

*And you are continuing to give talks?*

Yes, I have given a lot of talks around the world about my life with astrometry, and I wrote many reports about the history of astrometry. After my talk in Stuttgart in 2019, when we received the instrumentation prize from the German Astronomical Society, I spoke in Paris and in Berlin. I've given other more recent talks online, in 2020 first to an audience in Potsdam, then to Perth in Australia invited by my colleague Andreas Wicenec from the Tycho and Tycho-2 projects when he worked in Tübingen, and to South Africa invited by Mattia Vaccari who was my student in Copenhagen in the 1990s, - Mattia came from Padua and studied how to map galaxies with Gaia data.

## Launch and after

*You'd flown out, at ESA's invitation as a VIP, to see the launch of Hipparcos by Ariane 4 from French Guiana in 1989.*

*With all of your contributions to the developments underpinning its successor, Gaia,*

*I was greatly disappointed that ESA did not recognise your central contributions to this next step when deciding on the attendees at its launch from French Guiana in December 2013.*

*I suppose that there was little `corporate memory' of the 20-year journey in making Gaia happen.*

*But you did see the launch relayed live in Copenhagen; tell us about your memories of this great day?*

No, I was not invited to see the launch of Gaia, and of course I was also disappointed.

But to my great delight, however, the day of Gaia launch came to please me very much in Copenhagen. My colleagues, especially Uffe Gråe Jørgensen, had arranged a meeting for the occasion at the Niels Bohr Institute (where I had my office) with invited guests, talks and transmission of the launch. I had invited several guests, my wife Aase was there, and my admired tutor from the 1950s, Professor Peter Naur, who I talked about in my first interview with you. After a short talk at the meeting I went with Peter Naur in a taxi to the TV studio where we both appeared for interview during the launch. You cannot imagine how happy I was to see him enjoying this event with me, his student from sixty years ago!

That must have been a remarkable and rather strange event with your tutor again after 60 years!! Did he show any particular surprize or pride in seeing what you had achieved?

Yes, I could see it and he said it is unbelievable what you have achieved Høg (Naur spoke Danish: *Det er utroligt, hvad du har nået Høg*). And I should add that we had frequent contact with each other on the telephone, until he died a few years later.

*So, Gaia was launched in 2013, and operations started in mid-2014, after a 6-month commissioning phase.*

*Our original target mission duration was for 5 years, but with consumables on-board for 10 years,*

*and thanks to the superb mission operations team in ESOC, and the science operations team in ESAC, Gaia has now acquired more than 7 years of high-quality data*

*with three Data Releases issued so far - in 2016, 2018, 2020 and the next one has just been announced for the 13 of June.*

*Erik, you are now 89, and 90 in June I think, and as we've said, you are not involved in the analysis of the data coming from Gaia*

*but I know that you are watching the many teams, and the huge data processing task involved.*

*Tell us something about your perspectives on this enormously successful mission and some of the scientific results that you have seen reported.*

It is overwhelming to see what has come out, and that the data reduction and everything has been able to follow up. Very good management, very good people. I just listened to your interview with Anthony Brown, leader of the data reduction consortium, it touched me deeply, almost to tears. How lucky I am to live long enough to see all this myself.

tm=39:10

*Yes, I feel very similar Erik, it is a wonderful and dedicated team, more than 400 people working hard to get these wonderful results out.*

*As I said at the end of our second interview…*

*in one lifetime you have helped take astrometry from the confines of the Earth*

*out into space with two highly successful space missions.*

*But your contributions haven't stopped with Gaia…*

*Tell us what you've been doing to help prepare for the future?*

In May 2013 I heard from Claus Fabricius (in Barcelona) about ESA's call for new large missions, just 11 days before deadline. The call was a surprise, but It was immediately clear to me that it should be a Gaia successor in 20 years. When writing the proposal I corresponded with colleagues, some urged me to submit it, a few others said I should not submit because it was not ripe and because such a proposal should be submitted by a committee, not by a single person. But I had only a few days to write and to decide. I submitted the proposal in time, which was 6 months before Gaia was launched.

I was sure the mission should use filter photometry with the same high angular resolution of 0.12 arcsec as the astrometry for the benefit of narrow double stars and dense fields. It should not be prism photometry as in Gaia with a resolution about one arcsec. The prism photometry of Gaia has its advantages, but there is no reason to repeat it. Likewise, the medium-dispersion spectrograph could be omitted since there are no strong scientific reasons to repeat these observations, and cost could be saved.

Please tell us about the scientific ideas behind this idea of a second Gaia ?

The idea was that astrometric observations of the Gaia stars after 20 years would give ten times more accurate proper motions from positions at the two epochs than from a single mission. The benefit for kinematic studies would be immense for more distant stars, and also for stars close-by due to better accuracy, and for maintaining the celestial reference system. Exoplanets with long Jupiter-like periods could be detected and the whole population of such systems in our neighbourhood could be discovered. During the following years I wrote a long report with all the scientific cases while corresponding with many colleagues and I presented the ideas at many meetings. I had response from at least a hundred competent people in oral discussion or correspondence.

tm=42:20

*But you eventually discarded the original idea of a second Gaia,*

*focusing instead on observations in the infrared, as the Japanese JASMINE project were considering?*

NO, NO Michael, nothing was discarded, we developed *a new capability* as I shall now explain.

With Jens Knude and other colleagues in Copenhagen I began in 2014 to look at the possibility to extend the wavelength range into the near infrared, beyond the 1000 nm possible with the classical CCDs. The benefit would be the observation of redder stars i.e. cooler stars or stars imbedded in nebulosity.

In 2015 a colleague at the Lund Observatory, David Hobbs, joined me with much energy as he still does and we focused on GaiaNIR, a successor in 20 years with an Near-InfraRed capability. A technical study was conducted by ESA in 2017. Especially the NIR detector is critical since we need TDI, time-delayed integration, which was easy to do in the visual with CCDs in Gaia but is difficult in NIR. The vast majority of the stars observed by Gaia shall be observed again by the successor since a good sensitivity at less than 1000 nm shall be maintained. The positions at two epochs will be obtained for them and thus the highly accurate proper motions.

An industrial study of NIR detection is making progress. The new mission will be able to peer into the obscured regions of the Galaxy and measure up to 10 or 12 billion new objects, in addition to two billion Gaia stars, and reveal many new sciences in the process. ESA has now ranked the development of this mission so high that a launch about 2045 is quite probable, see [3].

tm=44:40

*You mentioned 10 billion stars? May I quote some words from a decade ago from someone who I described at the time as an undisputed authority in astrometry, this person said that `The Gaia astrometric survey of a thousand million stars cannot be surpassed in completeness and accuracy within the next forty or fifty years.' Do you recognise those words!!!*

Yes Michael, of course I recognise these words, I believe I wrote them myself about 2005 and I am glad to hear them again from you. I knew how long it would take to get a new large mission approved and developed, and we can expect the first results from GaiaNIR in 2050 if all goes well - which is rather within the time frame I mentioned. The mission will supersede Gaia in two respects: The infrared capability and the 10 billion stars.

I am happy that I have been able myself to help bringing this bright future within reach. It is like a dream, but it seems to be true! It has happened so soon because the proposal with a realistic design was submitted in May 2013 even before Gaia had been launched and work was started on the project. Then two years later, David Hobbs became active for the Gaia successor. David has worked on Gaia since 2007 and he brought great scientific and technical insight and he has made full use of all the possibilities offered by ESA.

*As I mentioned before, you've received some welcome recognition for your work over the years;*

*And in particular, in 2019, you received the Instrumentation Prize of the German Astronomical Society,*

*it was jointly with Lennart Lindegren and me for our contributions to Gaia, and in your case, emphasising your lifelong work in astrometry.*

I was very happy about this recognition from the German Astronomical Society, and especially that I received it together with you, Michael and Lennart. We three have enjoyed a wonderful collaboration for so many years, with Lennart it began in 1973 almost 50 years ago, and with you in 1981.

Let me also tell this Michael: When the European Astronomical Society announced its new prize in 2008, the Tycho Brahe Medal, I thought secretly, that I might get it! But when you then finally got it in 2011 I thought: *that is perfectly right*, Michael deserves it for his brilliant leadership of the Hipparcos and Gaia Teams since we four consortia leaders of Hipparcos had already received the Director of Science Medal in 1999.

tm=48:00

*You have played a leading role in the move of astrometry to space, and inspired many others,*

*But there is still much to do for the generations that will follow?*

Let me recall what we wrote in the proposal for the ROEMER mission 29 years ago: "*The mission is not primarily aimed at solving some particular problem, but rather to explore the rich complexity of Nature in order to find new patterns, anomalies and even contradictions to existing concepts*". This is also the spirit of the proposed GaiaNIR mission, and this vision has become true in a spectacular way for the Gaia mission when we see [4] that the number of scientific papers per year has been larger since 2019 than for all the other ESA missions taken together, and only based on the first 22 (DR1 and DR2) months of Gaia observations. Imagine how it will look after 10 years=120 months! - You will see that Michael.

## Closing off

*This has only touched on a few aspects of the Gaia mission and beyond, in particular…*

*…. some of the main points seen through the eyes of one of space astrometry's leading scientists*

*and it's been interesting to hear some of your thoughts as you look back over this important period*

*Erik Høg: thank you very much for joining me again today.*

Thank you for inviting me, and for your perfect and very kind guidance.

Allow me now, Michael, to leave astronomy for a moment and speak about my concern for our civilization which is shared by many people. NOT the planet Earth is in danger as we often hear in discussion of climate and global warming, because plants, animals and the human race will of course survive and adapt to the changing conditions. It is *"only"* human civilization that will suffer increasingly. The living conditions have been improving for a large fraction of people on Earth during the past centuries, but this will not continue for very long for many reasons: increasing pollution of atmosphere, soil and oceans, people smuggling, international crime, and because our political structures are unable to fundamentally reverse this trend.

But 500 years ago, Martin Luther spoke a good motto for everyone:

*"Even if I know that the world will perish tomorrow, I will plant a tree today."*

*Well, I think that astronomers - like us - can offer their own very unique perspective on this problem,*

*being so conscious of the size and emptiness of our Galaxy and beyond*

*and the fact that we do appear, so far at least, entirely alone in its vastness*

*Erik, thank you very much again.*

tm=51:30

## References

[1] Høg E. 2011, **Astrometry history: Roemer and Gaia**
26pp at: http://arxiv.org/abs/1105.0879

[2] Høg E. 2014, **Interferometry from Space: A Great Dream**

In: Asian Journal of Physics Vol. 23, Nos 1 & 2 (2014),

Special Issue on History of Physics & Astronomy, Guest Editor: Virginia Trimble.

http://arxiv.org/abs/1408.4668

[3] Høg E. 2021, **The Gaia Successor in 2021.**
3 pp at: http://arxiv.org/abs/2107.07177 [astro-ph.IM]
http://www.astro.ku.dk/~erik/xx/GaiaSuccessor2021.pdf

[4] ESA 2022, **Refereed publications from ESA missions.**

1 p at: http://www.astro.ku.dk/~erik/xx/GaiaPublications.pdf

Further references about the history of astronomy related to space astrometry in the early 1970s and about Erik Høg are given at the first interview. Most papers by Høg may be found on ADS by searching for "Høg, E." or "Hoeg, E.".